\documentclass[useAMS,usenatbib]{mn2e}
\usepackage{graphicx}
\usepackage{amssymb,amsmath}
\title[{\sc PynPoint}]{{\sc PynPoint}: An Image Processing Package \\for Finding Exoplanets}
\author[A. Amara \& S. P. Quanz]{Adam Amara\thanks{adam.amara@phys.ethz.ch} and Sascha P. Quanz\thanks{sascha.quanz@astro.phys.ethz.ch}\\
Institute for Astronomy, ETH Zurich, Zurich 8093, Switzerland}

\begin{document}

\date{}

\pagerange{\pageref{firstpage}--\pageref{lastpage}} \pubyear{2012}

\maketitle

\label{firstpage}

\begin{abstract}
We present the scientific performance results of {\sc PynPoint}, our Python-based software package that uses principle component analysis to detect and estimate the flux of exoplanets in two dimensional imaging data. Recent advances in adaptive optics and imaging technology at visible and infrared wavelengths have opened the door to direct detections of planetary companions to nearby stars, but image processing techniques have yet to be optimized. We show that the performance of our approach gives a marked improvement over what is presently possible using existing 
methods such as {\sc LOCI}. To test our approach, we use real angular differential imaging (ADI) data taken with the adaptive optics assisted high resolution near-infrared camera NACO at the VLT. These data were taken during the commissioning of the apodising phase plate (APP) coronagraph. By inserting simulated planets into these data, we test the performance of our method as a function of planet brightness for different positions on the image. We find that in all cases {\sc PynPoint} has a detection threshold that is superior to that given by our {\sc LOCI} analysis when assessed in a common statistical framework. We obtain our best improvements for smaller inner working angles (IWA). For an IWA of $\sim$0.29$''$ we find that we achieve a detection sensitivity that is a factor of 5 better than {\sc LOCI}.  We also investigate our ability to correctly measure the flux of planets. Again, we find improvements over {\sc LOCI}, with {\sc PynPoint} giving more stable results. Finally, we apply our package to a non-APP dataset of the exoplanet $\beta$ Pictoris b and reveal the planet with high signal-to-noise. This confirms that {\sc PynPoint} can potentially be applied with high fidelity to a wide range of high-contrast imaging datasets. 

\end{abstract}


\section{Introduction}

The field of exoplanet research has undergone rapid growth with a variety of methods having been developed to detect and charactere the properties of planets orbiting stars other than our Sun. Among these, radial velocity \citep[e.g.,][]{mayor2011,cumming2008,howard2010} and transit \citep[e.g.,][]{batalha2012,borucki2010} methods have yielded the majority of the planets that have been discovered thus far. Another approach is to image exoplanets directly. The difficulty with this method is that the flux of the planet is substantially smaller than that coming from the star it orbits. Hence, a simple image of the system will be overwhelmingly dominated by the star. This, in combination with the small projected separation on sky between the planet and the host, makes this imaging approach to exoplanet detection difficult. However, there have been discoveries of a few remarkable exoplanets and exoplanet candidates: HR8799 bcde \citep{marois2008,marois2010}, $\beta$ Pic b \citep{lagrange2009a,lagrange2010}, 2MASS1207 b \citep{chauvin2005}, LkCA15 b \citep{kraus2012}, 1RXS1609 b \citep{lafreniere2008}, 2MASS044144 b \citep{todorov2010}.  Despite these successes a large number of dedicated planet search surveys have yielded null results  \citep[e.g.,][]{masciardi2005,biller2007,kasper2007,lafreniere2007,chauvin2010,heinze2010}. While one of the key conclusions from these surveys has been that massive gas giant planets are rare in orbits with separations larger than a few tens of AU, the innermost regions around the stars are still largely unexplored as we lack the required contrast. To improve this situation, a number of technical methods have been developed that allow us to effectively suppress the flux of the star and thereby form the desired image of the planet. These innovations have been in the hardware and techniques for acquiring the data, as well as the analysis methods for data processing. 

At the data acquisition stage, new developments include new coronagraph designs \citep[for recent overview see, e.g.,][]{guyon2006} to suppress the flux and halo from the star at small inner working angles (IWA) and improved observing strategies for differential techniques that allow for better modeling of the point spread function (PSF). Examples of these methods are simultaneous differential imaging \citep[SDI,][]{lenzen2004} and angular differential imaging \citep[ADI,][]{marois2006}. The common goal of these approaches is to control the stellar PSF so as to reduce `speckles', which are residual features in the (final) images that can mimic and/or overshadow a planet. 
Upcoming new dedicated instruments for planet finding combine several of these methods and contrast enhancing techniques with extreme adaptive optics (AO) systems, yielding unprecedented PSF stability and contrast performances \citep[e.g., SPHERE at the VLT, GPI at Gemini;][]{beuzit2006,macintosh2007}.

Along with these improvements, new data analysis methods have been studied. This is a very important part of the process since inefficient methods will not allow us to take full advantage of the new higher quality datasets that are becoming available. On the other hand, highly efficient software solutions allow us to reanalyse existing data to extract more information. A good example is the widely used {\sc LOCI} package \citep{lafreniere2007b}, which is used to model and subtract the stellar PSF. This method significantly improves the contrast performance in ADI datasets compared to the classical ADI data reduction approach and, in the meantime, has also been be adapted to effectively subtract the sky background emission in thermal infrared datasets \citep{galicher2011}.

Here, we present the results of a new data analysis and PSF subtraction software package - called {\sc PynPoint} - that  we have developed. Our method shows significant improvements to the contrast performance at very small IWAs in ADI datasets. Compared to existing methods, we gain up to a factor of five in sensitivity at separations of $\sim$0.29$''$. {\sc PynPoint} can be applied to a wide range of datasets, including any existing ADI dataset and does not require any special observing strategy or instrument setup. 

In section 2, we give a description of the dataset and simulations we used. In section 3, we give an overview of the main top-level steps used in {\sc PynPoint}. In sections 4 and 5, we show our performance in both planet detection and flux measurements. We compare our results with those we obtained using {\sc LOCI}.  In section 6, we apply {\sc PynPoint} to a high-contrast dataset of the exoplanet $\beta$ Pic b. Our conclusions and discussion on future prospects are presented in section 7.

\section{Description of Test Data}\label{testdata}

To test and validate {\sc PynPoint}, we  created a suite of simulations based on real ADI data. The dataset we used consisted of images taken with the adaptive optics (AO) assisted high resolution camera - NACO - at the VLT \citep{lenzen2003, rousset2003}. These images were taken during the commissioning of the apodising phase plate \citep[APP;][]{kenworthy2010,girard2010} coronagraph, which took place in April 2010. We observed the young, nearby debris disk host star HD115892 in pupil stabilised ADI mode in the NB4.05 filter and used a detector integration time (DIT) for individual exposures of 0.5 s. A complete description of the observations and a first data reduction and analysis (done with {\sc LOCI}) is presented in \cite{quanz2011}. Despite the fact that we did not detect any faint companions around HD115892, this dataset is well suited for our current purposes since the APP already provides a good contrast performance close to the star using current reduction techniques. The full test dataset consists of 10,800 individual exposures contained in 54 data cubes (all images from `hemisphere 1', see \citet{quanz2011}) which corresponds to a total on-source integration time of 1.5 hours. The field rotation over the full image stack was $\sim50^\circ$.

\begin{figure}
\begin{center}
\includegraphics[width=90mm]{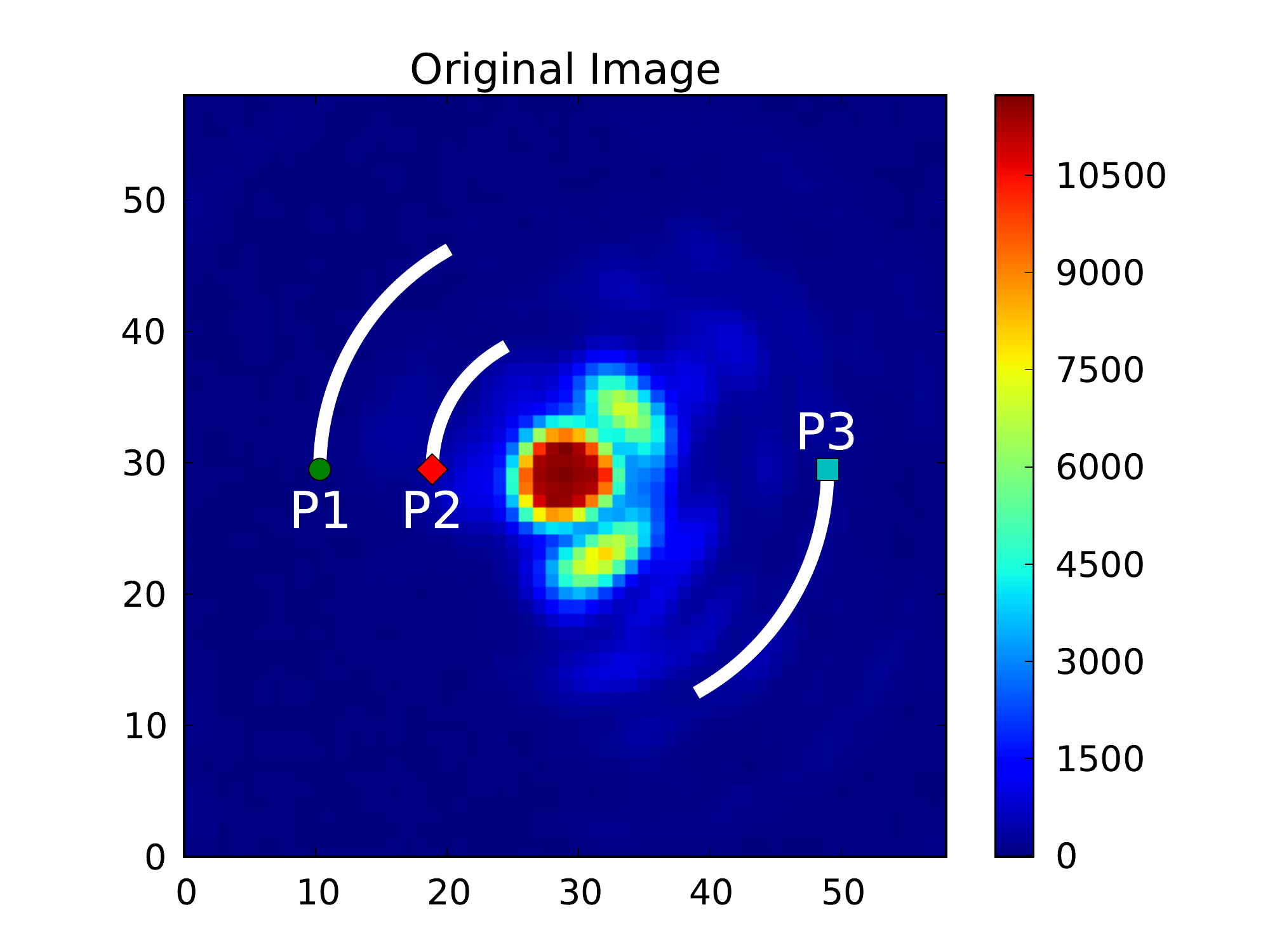}
\includegraphics[width=90mm]{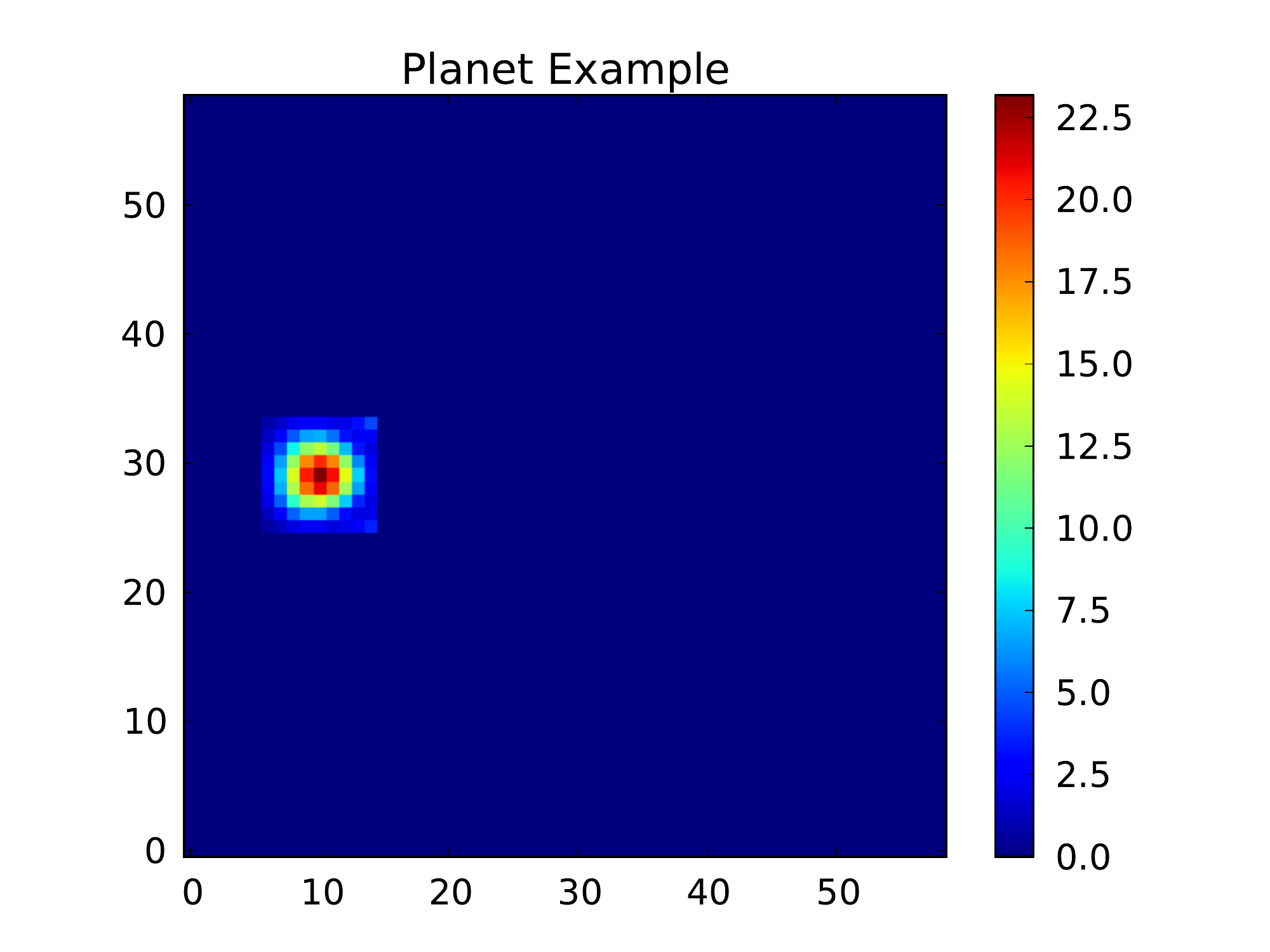}

\caption{The top panel shows an example of a PSF as observed in an individual frame. On the image we show the three positions of the planets that we used to test our performance. We will refer to these positions as P1(shown by the green circle symbol), P2 (red diamond) and P3 (turquoise/blue square). The symbols show the starting positions of each of the planets and the arcs show the track of each planet during the ADI observing sequence. Note that in our simulations only one planet is used at a given time. The lower panel shows an example of one of the planets that we added to our simulated images. This planet is the brightest example ($\Delta$ NB4.05 = 8 mag). For the faintest example, we used ($\Delta$ NB4.05 = 11 mag), i.e., $\sim$16 times fainter. The unit of the x and y axes in the images is pixels, while the flux levels are given in detector counts.}
\label{fig:input_images}
\end{center}
\end{figure}

The basic data reduction steps (background subtraction, bad pixel correction, and image alignment) are described in \citet{quanz2011} and the only additional step was to create postage stamp images from each exposure by cutting out the innermost 59x59 pixels centered on the star. From these postage stamps, we then created a suite of simulations by inserting fake planets to the individual raw images. Each of our simulations contains one planet at one of the three positions shown in Figure \ref{fig:input_images}. For the image of the planet, we used a scaled version of an unsaturated image of HD115892 \citep[ this image was created from averaging over a stack of 200 individual unsaturated frames; see also][]{quanz2011}. The full-width-half-maximum (FWHM) of the unsaturated PSF was $\sim$4.5 pixels. By varying the brightness of the planet, we can study the sensitivity of our method as a function of the magnitude contrast between the central star and the planet at a a given position.

\section{Overview}\label{pynpoint_overview}

{\sc PynPoint} is a Python-based software package that we have developed for processing high contrast imaging data of exoplanets. Once the data from an observation run has been reduced to a stack of postage stamps, four main steps are needed for our planet detection procedure. These are:
\begin{enumerate}
  \item  constructing a basis set for the analysis;
  \item  fitting the stellar PSF to the individual images;
  \item correcting for the PSF and central star; and
  \item averaging over an image stack.
\end{enumerate}
We have implemented methods in our package for dealing with these four steps. In so doing, we have made improvements in all the steps, but in particular our main gain comes from an improved treatment of the first step. Specifically, we have used a Principle Component Analysis (PCA) approach to empirically build the basis set from the data. In the following sections, we discuss each of these steps in more detail.

\subsection{Construct Basis Set}
\subsubsection{Overview of Basis Functions}

One of the key challenges in analysing high contrast imaging data is to correct for the flux of the central star. The star is effectively a point source, so its image will be that of the PSF of the respective telescope-instrument combination altered by atmospheric effects (for ground-based observations). This PSF can be modeled in a number of ways, and one effective approach is to decompose it into a set of basis functions so that the star image can be described through a linear combination such that 
\begin{equation}
I(\vec{x}) = \sum a_i \phi_i(\vec{x}),
\end{equation}
where $I(\vec{x})$ is the image of the PSF, $\phi(\vec{x})$ is a given basis and $a_i$ is the coefficient for each basis function. Often it is convenient to work with orthonormal basis functions, such that 
\begin{equation}
\int \phi_i(\vec{x}) \phi_j(\vec{x})  ~d\vec{x}=\delta_{ij},
\end{equation}
where $\delta_{ij}$ is the Kronecker delta. This is because it is computationally convenient to use such orthogonal basis functions, since it is easy to calculate the coefficients, $a_i$, for a given image $I(\vec{x})$, using
\begin{equation}
a_i = \int I(\vec{x})\phi_i(\vec{x})~d\vec{x}.
\label{eq:ai}
\end{equation}
While in this work we have focused on orthogonal complete basis sets, it should be noted that it is also possible to use over-complete basis sets, where the basis functions are not strictly orthogonal to each other. 

There are a number of popular basis set functions that have been widely used and studied. As an illustration, two such examples of basis set functions are: (i) Fourier -   decomposition into sine and cosine functions; and (ii) Shaplets -  a decomposition into Gaussian weighted Hermite polynomials \citep[for example see][]{2003MNRAS.338...35R}. Another approach, which is the one that we have adopted, is to empirically create a basis set from the data. Typically, this type of empirical approach leads to basis functions that are more efficient at expressing the underlying function than generic basis sets. Here, the efficiency typically refers to a measure of the number of coefficients that would be needed so that residuals between a model and the original images are lower than a given threshold. Efficient basis functions will require a smaller number of coefficients than non-efficient ones. Since many of the basis sets that we would consider using are complete (or over-complete), meaning that they can be used to describe any function when an infinite series is used, we might think that we are free to use any of them. However, since in practice we have to contend with noise, which effectively truncates the series, the choice of basis sets becomes important \citep[for an example of the impact of basis set choice, see][]{2009MNRAS.398.2134K}. In this case, it is better to use efficient basis functions. 

In weak gravitational lensing, the impact and importance of basis sets in modeling the PSF pattern on an image have been widely studied, due to the fact that a critical step in the weak lensing measurement process is to measure and correct for the PSF at the positions of galaxies \citep[for top-level overview of these processes, see][]{2009AnApS...3....6B,2011arXiv1109.3410A}. In particular, \cite{2007PASP..119.1403J} have compared shapelet, wavelet and PCA methods and shown that for their application the empirical PCA method gave the best results. Though different in detail, since in weak lensing we work with a spatially varying PSF while in exoplanet direct imaging the PSF varies primarily in time for ground-based AO-assisted instruments, the key aspects of the two problems share many similarities. For this reason, we have focused our initial studies, which we present here, on PCA basis functions.

\subsubsection{PCA Basis Sets}\label{pca_basis_set}
A brief discussion of the basic principles of PCA methods is given by \cite{2007PASP..119.1403J}. In this section, we give a short summary of the key calculations and outline some of the key issues that need to be considered. We calculated the principle components of our images using singular value decomposition \citep[SVD, see][]{1403886}. To do this, we constructed a two dimensional array ${\bf S}$, where each row $i$ of the array is a vectorised version of image $i$ of the stack. This means that ${\bf S}$ has dimensions ${\bf M}\times {\bf N}$, where ${\bf N}$ is the number of pixels in an image and ${\bf M}$ is the number of images in a stack. We calculate the SVD such that
\begin{equation}
{\bf S} = {\bf UWV^T},
\label{eq:SVD}
\end{equation}
where ${\bf W}$  is a diagonal matrix with positive (or zero) elements. In this decomposition, ${\bf V}$ is a matrix containing the PCA elements that form an orthogonal basis set (${\bf U}$ and ${\bf V}$ are column-orthogonal matricies).

\begin{figure*}
\begin{center}

\includegraphics[width=180mm]{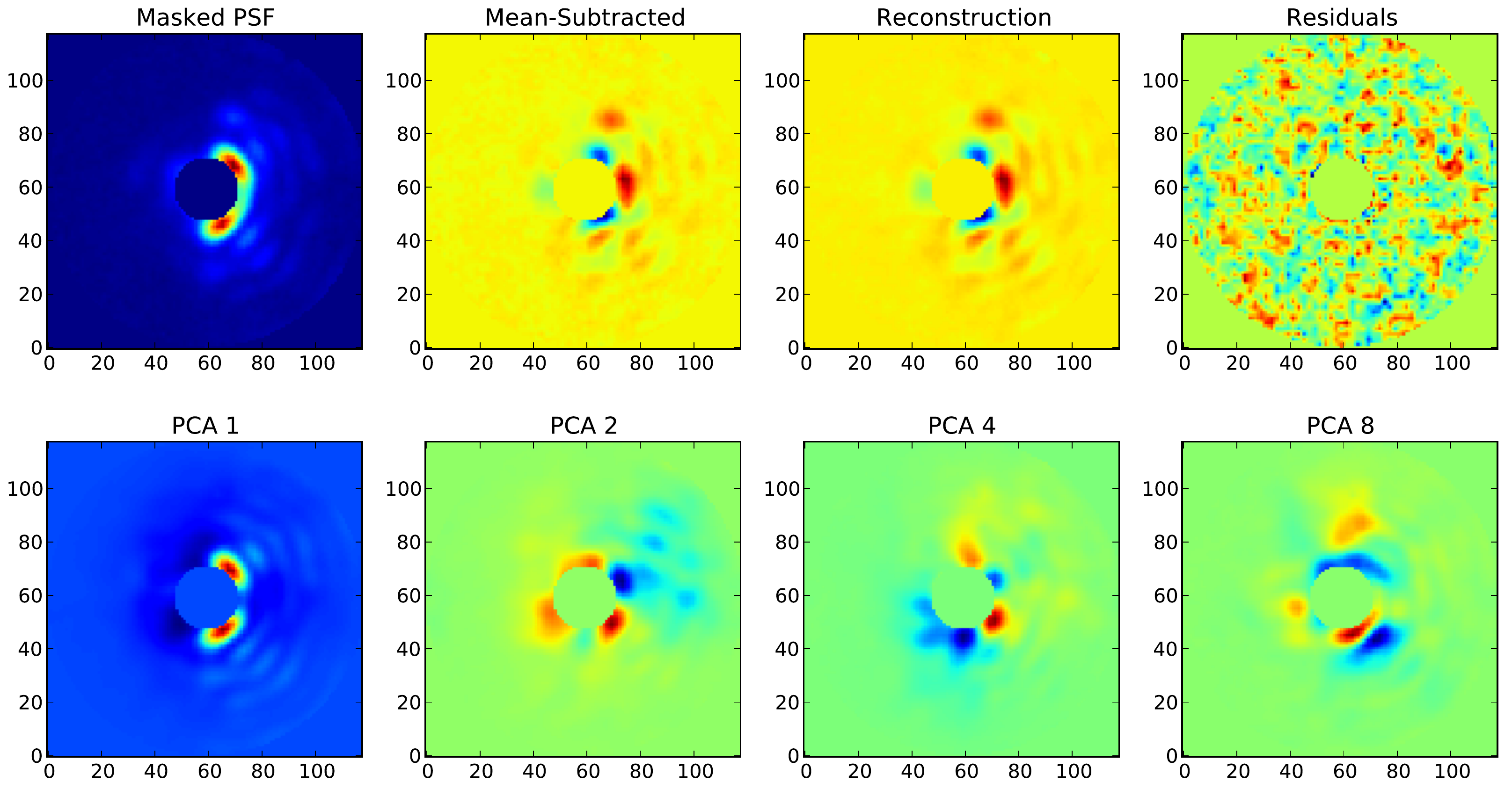}

\caption{The top panels show an example of our PSF fitting approach. From left to right, we show: (i) a masked version of a single image PSF where the centre and the corners of the image have been removed; (ii) the same PSF with the average over the stack removed; (iii) the reconstruction with 100 PCA coefficients; and (iv) the residuals between the mean subtracted PSF in the first panel and our reconstruction. In the bottom panels, we show four examples of our PCAs.}
\label{fig:PCA}
\end{center}
\end{figure*}

After running a series of tests, we were able to optimise our PCA decomposition approach by using a careful treatment of the data before applying the decomposition shown in equation \ref{eq:SVD}. Three important steps, which can be seen in the top panels of Figure \ref{fig:PCA}, are:

\begin{enumerate}
  \item  Subpixel Sampling. We have a large number of PSF realisations, with each one expected to have small offsets between the PSF centre (position of the star) and the pixel grid set by the detectors. This then allows us to reconstruct features over the stack at a resolution that is greater than that of any individual image. For instance, in the example shown in Figure \ref{fig:PCA} we have doubled the resolution of the original images to $118\times118$ pixels as compared with the input images (shown in Figure \ref{fig:input_images}) with a resolution of $59\times59$.
  \item Central Mask. We know that the central region dominates the flux of the image, and in our HD115892 dataset the core is even saturated \citep{quanz2011} and contains little useful information. We find that our analysis methods work best when this central region is masked. This is shown in the top left panel of Figure \ref{fig:PCA}.
  \item Mean Subtraction. When constructing the PCA basis functions, we have the option of subtracting the mean image, which is generated from the entire stack,  from each of the individual images before the decomposition. This should be a subtle effect, since if the stack has a mean image then this would likely be (or dominate) the first principle component. Hence, we could expect the step of removing the mean to have little effect. However, since the PCA basis functions are forced to be orthogonal to each other, including the mean as one of the PCAs forces the higher PCA functions to be orthogonal to it. By first subtracting the mean, this condition is removed and we find that the resulting PCA functions work better. An example of a single mean subtracted image is shown in one of the upper panels of Figure \ref{fig:PCA}.
\end{enumerate}

The bottom panels of Figure \ref{fig:PCA} show a subsample of the PCA functions for the example we have used in this study. Specifically, we show the PCA functions 1, 2, 4 and 8. In the upper center-right panel we show the reconstruction of a single mean-subtracted image using 100 PCA coefficients. In the top right panel, we show the residuals between the data and the model.

\subsubsection{Further Considerations}

One important consideration in empirically building the basis set is to decide which data to use. For instance, if we use the same data to construct the PCAs as what we will use to detect the planets, then there is a danger that the PCA will incorporate the planet and so will tend to remove it from the analysis. There are a number of ways that we could consider to overcome such a problem. For instance, the data could be divided into a training sub-set and an analysis sample. Alternatively, we could try to build a generic basis set for a particular instrument using, for instance, archive data of different stars. Since the training set would be independent of the data used to make the measurements, we reduce biases coming from the modeling. The disadvantage is that these PCA functions are likely be less efficient. 

A further consideration is that the matrix operations in equation \ref{eq:SVD} can be computer memory intensive when the matrix ${\bf S}$ becomes large. Ultimately, this will likely lead to a trade-off between the number of images in a stack and the sub-pixel resolution we wish to recover. These optimisations still require further refinement, but our finding in the work presented here suggests that using the same stack to construct the PCA as is used in the data analysis does not present a serious problem. We believe this is because 1) the planets are substantially fainter than the star, and 2) the planets ``rotate" around the star in our ADI observations.

\subsection{Fitting the PSF}\label{psf_fitting}

For each image in our stack we are able to calculate the PCA coefficients using equation \ref{eq:ai}. When using the whole image, the only variable that needs to be specified by the user is the number of PCA coefficients that should be used in the fit. For the case that we are studying here, we find that very good results are achieved for roughly 100 PCA coefficients per fit. Beyond this we find that the improvements in the residuals are marginal. This process is the simplest fitting procedure that we could implement. Nonetheless, we have also incorporated a fitting scheme that allows the user to specify a mask. The reason to do this is that in some cases, for instance when we want to measure the flux of the planet, it is useful to mask out the planet location and only use the data in the rest of the image to fit a PSF model. This then limits the extent to which the planet itself is included in the fitting and avoids overcorrection and reduces possible biases in flux measurement. The disadvantage is that the fit inside the masked region, by necessity, will not be as good. Therefore, random error will increase. Another complicating factor is that once a mask is introduced, the resulting basis functions (multiplied by the mask) are no longer orthogonal. This means that we can no longer find the linear coefficients using equation \ref{eq:ai}. However, since the masks that we typically introduce tend to be small, equation \ref{eq:ai} still gives a good first approximation to the best-fit coefficients.  After this, we then refine the coefficients, using a minimisation algorithm, to minimise the residuals,
\begin{equation}
R = \sum (I - I_m)^2\times M,
\end{equation}
where $I_m$ is the model PSF images for that frame, M is the mask image that blocks out the planet position and the sum is performed over all pixels in the image. This methods works because many of the features that a PSF has within the masked region are correlated to its properties in the unmasked regions. Given a good model for a PSF and a good fit in the unmasked region, we should be able to perform a reasonable reconstruction of the PSF where the mask sits. This method of refining the linear coefficients can be further optimised by minimising the weighted residuals, $R_w$, given by
\begin{equation}
R_w = \sum (I - I_m)^2\times M\times W,
\end{equation}
where $W$ is a weighting functions. In the work presented here, we used a Gaussian weighting function centered on the planet positions. This forces the coefficients to those that give a model that is in good agreement with the data in the region close to the mask (and planet) position. These features tend to be more strongly correlated with the features inside the mask, thus giving slightly better fits. The specific optimal weighting scheme is likely to depend on the details arising from, for example, the particular instrument and  telescope being used.

\subsection{PSF Correction and Averaging Over Stack}\label{stack_average}

For each image in the stack, we want to remove the flux from the star and to leave the image of the planet. The star light can be well modeled as a convolution between a delta function and the PSF of the observations for that frame. Since the intrinsic image is simple, removing the star light in the convolved image corresponds to a simple subtraction of a PSF model from the image. This should then leave an image of the planet convolved with the PSF. Since the model that we fit does not perfectly reproduce a given PSF, the planet image will still be subdominant to the noise in the residuals. We still need to average over the stack of images to boost the signal-to-noise of the planet. To do this, we derotate the images in the stack to the same on-sky orientation based on the changes in the parallactic angle of each frame. We then average over the stack. As well as the mean flux at each point, we also compute the variance, which gives us a measure of the noise in the temporal direction at each point in the averaged image.

Our final average image should then give us a good representation of the planet convolved with a PSF that is effectively an average PSF over all the data. On top of this, there remains considerable small-scale spatial noise that can be seen as pixel-to-pixel variations. To deal with this and to boost the detectability of the planet, we suggest two approaches to smoothing out these small-scale features. The first is to deconvolve the image with a crude estimate of the average PSF. This will have the effect of pulling all the flux of the planet into one pixel. The other approach, which we have adopted in the current work due to its simplicity, is to use a matched filtering scheme. In this approach, the image is convolved with an estimate of the PSF, and the value of the convolved image at the planet position is then a measure of the total flux of the planet.

\section{Planet Detection}\label{planet_detection}
We have investigated the efficiency of planet detection for the data presented in section~\ref{testdata}. We do this by creating a suit of simulations with planets of varying brightness at the three positions shown in Figure \ref{fig:input_images}. We refer to these positions as P1, P2 and P3, and they have star planet separations of $0.52''$, $0.29''$ and $0.52''$, respectively. It is worth noting that P1 and P2  lie in the high-contrast hemisphere of the APP PSF where the diffraction rings are suppressed, while P3 lies in the bad hemisphere where the diffraction rings are enhanced \citep[see][]{kenworthy2010}.
We varied the planets' brightness to achieve a star-planet contrast between 8.0 and 11.0 mag with incremental steps of 0.5 mag and analysed this suit with both {\sc LOCI} and the {\sc PynPoint} procedure outlined in section \ref{pynpoint_overview}. 

\subsection{Data Reduction Using {\sc LOCI}}

Before running the {\sc LOCI} package, we median-combined every consecutive 10 exposures in each cube, which resulted in 19 averaged frames per cube or 1026 frames in total for the 54 cubes we observed. Given the short integration time, the field rotation between the individual exposures was negligible, and thus the median-combined frames did not suffer significant smearing effects of the simulated planets. We then applied the {\sc LOCI} package on the 1026 frames as described in \citet{lafreniere2007b}, using the following {\sc LOCI} coefficients (same naming convention as in the original paper): FWHM=4.5 px, $N_\delta$=0.75, $dr$=5 and $N_A$=300. The PSF-subtracted frames were then de-rotated and median-combined to create the final {\sc LOCI}-reduced images.

\subsection{{\sc PynPoint} vs. {\sc LOCI}}\label{pynpoint_vs_loci}
For the {\sc PynPoint} data reduction, a particular detail to note is that in fitting the PSF model we have not used the masking procedure discussed in section~\ref{psf_fitting}. This is because performing a blind search over the image with a mask is computationally very intensive, and our current findings show that even without this step we are already able to reach results that outperform what {\sc LOCI} is able to achieve. However, we return to the masking procedure in section~\ref{planet_flux} when we focus the discussion on measuring the flux levels of planet candidates. We also note that we were able to perform the {\sc PynPoint} analysis using the full stack of 10,800 individual exposures. 

\begin{figure}
\begin{center}
\includegraphics[width=	85mm]{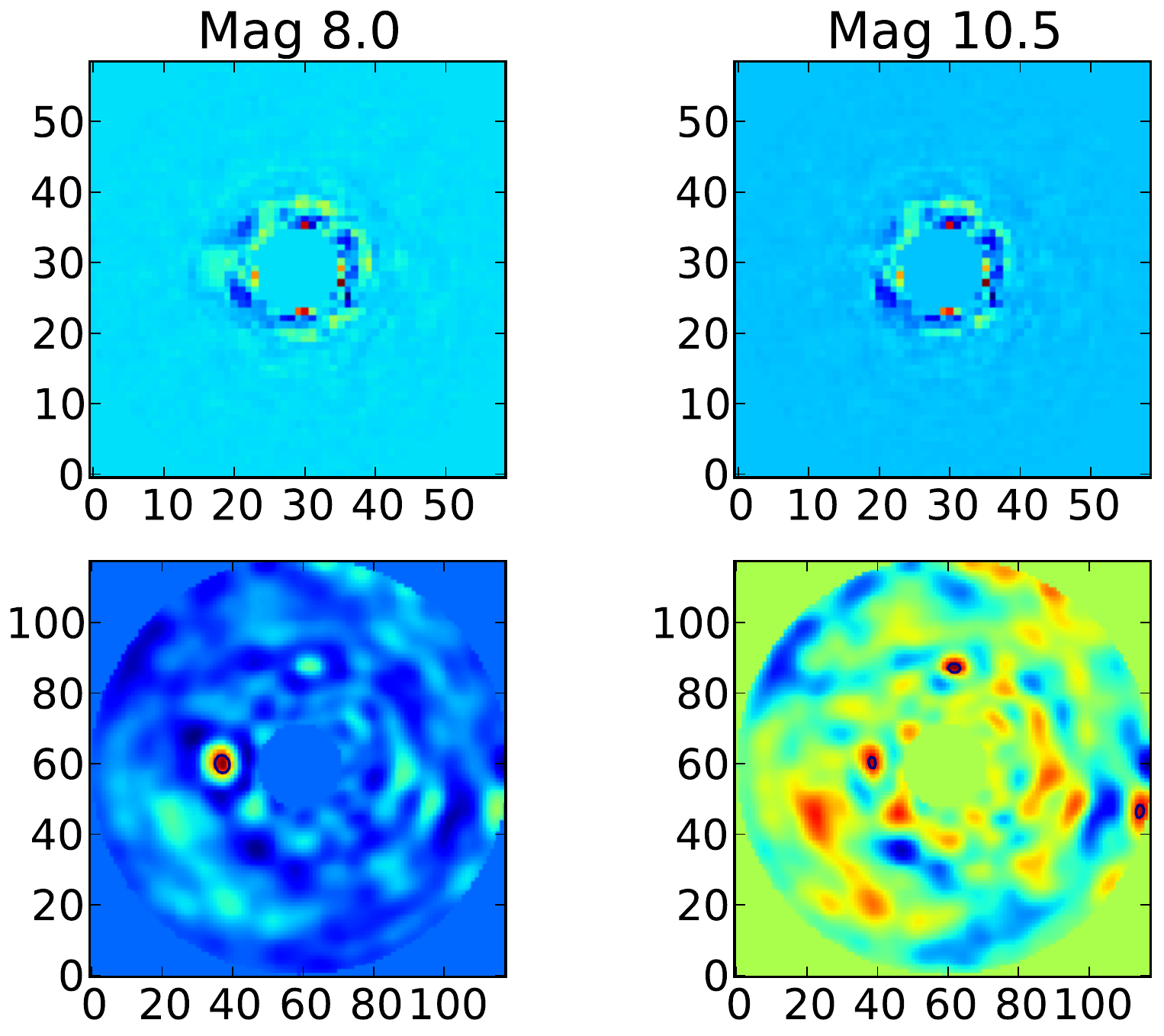}
\caption{Images used to detect the planets in position $P2$. The left panels show the cases for a planet with magnitude contrast of 8.0, and the right panels show results for the planet with a contrast of 10.5. The top two panels show results for {\sc LOCI}, while the lower panels are the results {\bf $I_D$} from {\sc PynPoint}.
}
\label{fig:residual_images}
\end{center}
\end{figure}

In Figure \ref{fig:residual_images}, we show {\sc LOCI} and {\sc PynPoint} results for the two cases where the contrasts between the star and the planet are 8.0 mag and 10.5 mag. These results all use planets located at position $P2$, which is the inner most configuration separated $0.29''$ from the star. 

The {\sc LOCI} data reduction is explained in the previous section. To generate the {\sc PynPoint} results, we have created an image that we have specifically tailored to object detection. We do this by creating a detection image, $I_D$, which we produce by dividing the average of the residual images by the root of the variance image,

\begin{equation}
I_D = {I_{res}} / \sqrt{I_{var}}. 
\end{equation}

We do this  because $I_{var}$ is a measure of the noise in each pixel and performing this operation produces an image with a more even noise distribution. This does suppress the signal in high noise regions, as it should, but in particular for a signal-to-noise analysis it does allow us to work more directly with the entire image rather than being totally confounded to working in rings around the star as is typically done. This is important since the {\sc PynPoint} detection images have been convolved in the final step, as outlined in section~\ref{stack_average}, to allow us to effectively match filter. In the example shown in Figure \ref{fig:residual_images}, we used a Gaussian of width of $\sigma = 0.027''$ (1 pixel in original image resolution) as a convolution kernel. For the {\sc PynPoint} images we have also applied an automatic detection routine, the results of which are shown by the contours. We see that for the bright planet (contrast of 8.0 magnitudes), the automatic algorithm easily finds the input planet. For the fainter planet, we see that the planet signal becomes comparable to some of the highest noise peaks, and in the image we see that the planet is one of the three brightest peaks. By eye, we see in all cases that there seems to be a substantial improvement in detectability for the {\sc PynPoint} images as compared with the {\sc LOCI} results.

\begin{figure}
\begin{center}
\includegraphics[width=90mm]{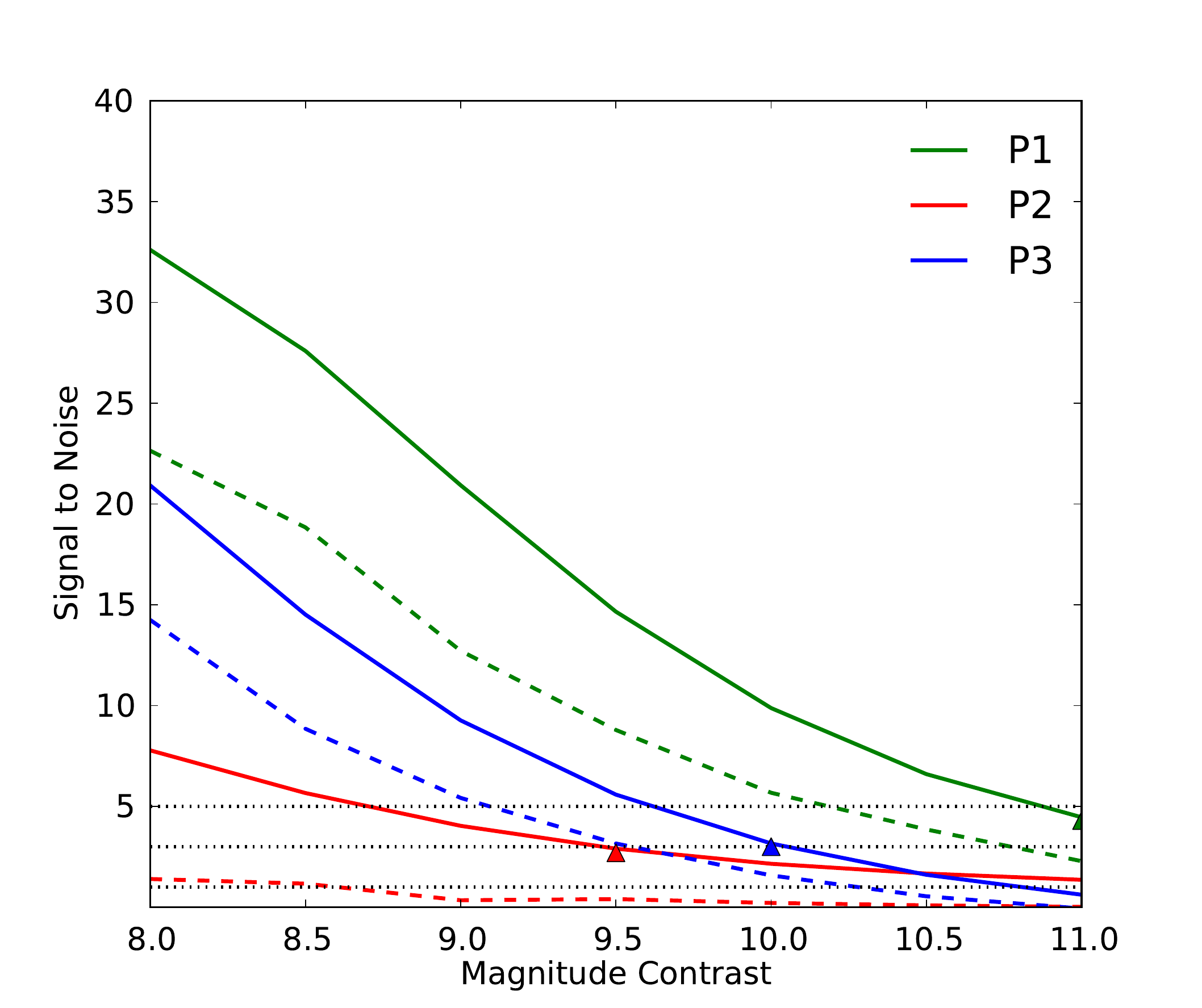}

\caption{Detection signal-to-noise (defined in section~\ref{pynpoint_vs_loci}) as a function of the contrast difference between the central star and the planet. The solid curves show results for {\sc PynPoint}, and the dashed curves show results for {\sc LOCI}. The colours correspond to the three planet positions that we have used in this study. The horizontal dotted lines show the signal-to-noise thresholds of 5, 3, and 1.}
\label{fig:detec_sn}
\end{center}
\end{figure}

In Figure \ref{fig:detec_sn}, we quantify the improved performance of our {\sc PynPoint} software package by showing the detection signal-to-noise as a function of planet magnitude for the three positions we are considering. In all cases we measure the signal from the peak value of the Gaussian convolved image at the planet position (for example, the peak value in the lower left panel of Figure \ref{fig:residual_images}). For the noise, we measure the standard deviation along the circumference of a circle that is centred on the star and has a radius that passes the center of the planet. The solid curves in Figure \ref{fig:detec_sn} show the {\sc PynPoint} results, the dashed curves show the {\sc LOCI} results and the different colours correspond to position $P1$ (red), $P2$ (blue) and $P3$ (green). The triangular symbols in the figure show the points where the planet peak is no longer the brightest peak over the entire image. We were only able to produce these results, shown with the symbols, for the {\sc PynPoint} images because the {\sc LOCI} images were dominated by the noise in the central regions and there were no {\sc LOCI} cases where the planet is the most dominant peak. Figure~\ref{fig:detec_sn} also shows 1$\sigma$, 3$\sigma$ and 5$\sigma$ in horizontal dotted lines.

For the two outer planet positions ($P1$ and $P3$), we find that using {\sc PynPoint} leads to a 50\% boost in the signal-to-noise, which is a substantial improvement. Seen another way, for a fixed signal-to-noise threshold we see that when the data is analysed using {\sc PynPoint} we are able to detect planets that are between 0.5 to 1.0 magnitude fainter than what is possible with {\sc LOCI}. For the close-in planet position $P2$ at a separation of $\sim0.29''$, the improvement is even more striking. At this position, we see that the {\sc PynPoint} signal-to-noise is boosted by roughly a factor of five. This is a very significant improvement. In fact, in the {\sc LOCI} analysis the planet at this position cannot be detected at any of the brightnesses we consider. However, using {\sc PynPoint} we are able to detect the planet at $5\sigma$ down to a magnitude of 8.6 and at $3\sigma$ down to magnitude 9.5. The importance of this improvement is even more significant when we consider the fact that the regions closest to the star, where {\sc PynPoint} performs best as compared with {\sc LOCI}, are scientifically the most interesting ones. These regions, which are as close to the star as possible, are the most important ones in terms of planet searches since direct imaging surveys have already demonstrated that giant planets are rare at large orbital separations \citep[beyond a few tens of AU, see, e.g.,][]{lafreniere2007,chauvin2010}, but the parameter space closer to the star is still largely  unconstrained.

\subsection{Potential Future Improvements}
Once a possible planet detection is found, it then makes sense to reanalyse the data by placing a mask over the target position and using the method that we outlined in section~\ref{psf_fitting}. Though computationally more intensive, this step is feasible, since it only needs to be performed at specific locations. With this in mind, it is interesting to consider what an optimal detection strategy might be. For instance, we could imagine that in a two-step process it would make sense to have a more relaxed, but clearly defined, criterion for the first detection (i.e. the step described in this section), such as the detection of all peaks higher than $3\sigma$ rather than $5\sigma$, which is more typical. All these candidates can then be studied in more detail using the masking procedure, as we do in section~\ref{planet_flux}, at which point more aggressive cuts could be applied. We have not yet fully investigated this optimisation, but it certainly warrants further systematic work in the future.

\begin{figure}
\begin{center}
\includegraphics[width=90mm]{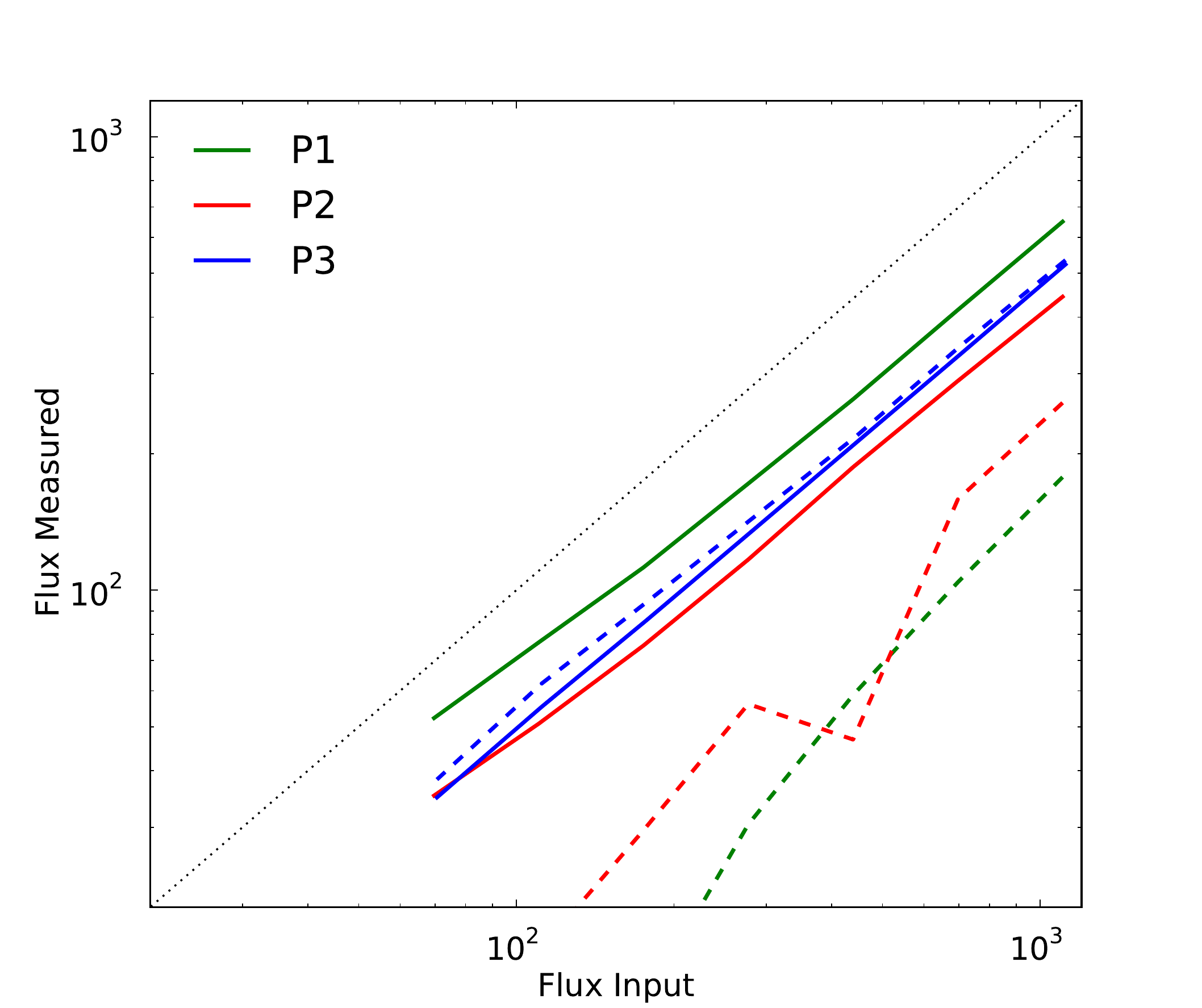}
\caption{Measured flux as a function of intrinsic planet flux. The black line shows the case where the measured flux is the same as that of the planet; the solid coloured curves show the results from {\sc PynPoint}, while the dashed curves show the {\sc LOCI} results. The colour scheme is the same as those used in Figure \ref{fig:detec_sn}.}
\label{fig:fluxin_fluxout}
\end{center}
\end{figure}

\section{Measuring the Planet Flux}\label{planet_flux}

It is well known that {\sc LOCI} introduces flux losses to companions \citep{lafreniere2007b}. These flux losses depend on the choice of {\sc LOCI} parameters and also on the angular separation and the brightness of the companion itself, thus making an accurate analytical correction difficult. Hence, in our next tests we studied the accuracy of {\sc LOCI} and {\sc PynPoint} at recovering the flux of the planet once its position is given. 

In the {\sc PynPoint} case, this means that we want  to use the residual images $I_{res}$ and not the detection images $I_D$, and we also use the planet masking procedure when modeling the PSF to limit possible biases. The results are shown in Figure \ref{fig:fluxin_fluxout}. The colour scheme and line styles are the same as those in Figure~\ref{fig:detec_sn}. The x-axis of this figure shows the flux inside an aperture with a radius of $0.064''$ ($\sim$2.4 pixels in original image resolution) centered on the input planet image (e.g., see lower panel of Figure~\ref{fig:input_images}). The y-axis shows the flux that is measured using the same procedure on the {\sc LOCI} and {\sc PynPoint} images. The diagonal dotted black line shows what we would expect in an idealised case where the measured flux is the same as the input flux. In general, we see that in all cases there is a bias between the input flux and measured flux. The important thing to note here is that the three {\sc PynPoint} cases give reasonably stable off-sets, while the {\sc LOCI} results show a large range in their performance. In the worse cases of $P1$ and $P2$, {\sc LOCI} underestimates the planet close to a factor of 5. While the two worse cases for {\sc PynPoint} underestimate the flux by a factor of 2. This combined with the better stability of the {\sc PynPoint} results means that {\sc PynPoint} is more likely that it can be calibrated. This leads to another example of better performance of our method. We believe that further improvements are possible through continued work and optimisation.

\section{Applied Example --- The Exoplanet $\beta$ Pictoris b}
\begin{figure}
\begin{center}
\includegraphics[width=85mm]{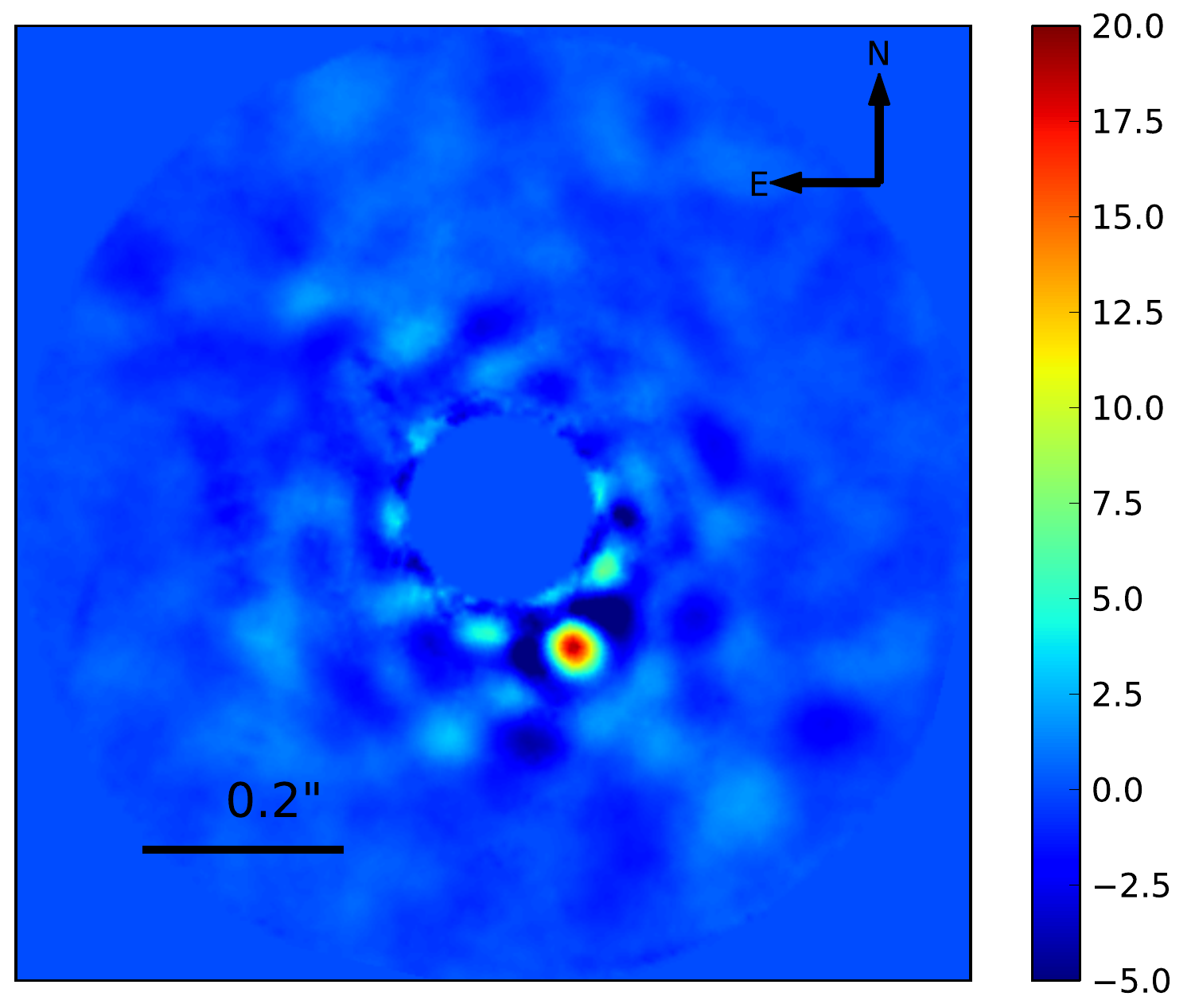}
\caption{The exoplanet $\beta$ Pictoris b observed in December 2009 in the $L'$ filter as revealed with {\sc PynPoint} (see text). The image has been normalised by the standard deviation ($\sigma$) in a ring at the planet radius. The $\pm 15^\circ$ around the planet has not been included in the calculation of $\sigma$.}
\label{fig:beta_pic_b}
\end{center}
\end{figure}

To further test and validate {\sc PynPoint}, we downloaded a high-contrast imaging dataset for the exoplanet $\beta$ Pictoris b. The data are publicly available from the ESO archive and were initially used to confirm the existence of this young, massive planet orbiting the nearby A-type star $\beta$ Pic \citep{lagrange2010}. The dataset we used was taken on 2009 December 26 with VLT/NACO in the $L'$ filter in ADI mode. It consists of 80 data cubes, each containing 300 individual exposures with a detector integration time of 0.2 s. The total field rotation amounted to $\sim$44$^\circ$ on sky. Again, the basic data reduction steps (sky subtraction, bad pixel corrections and alignment of images) were done as explained in \citet{quanz2011}. The final postage stamps we used for our {\sc PynPoint} analysis were 73$\times$73 pixels in the original images size, but as explained in section~\ref{pca_basis_set}, we doubled the resolution of the images (in this case to 146$\times$146 pixels). Our {\sc PynPoint} analysis effectively has one significant free parameter that is set by the user. This is the number of PCA coefficients that are used to fit each image. However, we find that our results are fairly stable over a wide range of parameter choices. For $\beta$ Pictoris b we are consistently able to detect the planet with a signal-to-noise in the range of 15 to 20 when changing this parameter by a factor of 3 (between 20 and 60). We note that since the PSF of this non-coronagraphic dataset is less complex than that of the APP, fewer coefficients are needed than in the test datasets in the previous sections. For the results shown in Figure~\ref{fig:beta_pic_b}, we used 40 PCA coefficients. We see that the exoplanet $\beta$ Pictoris b is by far the strongest signal in the image and is clearly detected with a signal-to-noise of $\sim$20 south-west of the star at a position angle of $\sim$209$^\circ$. 

\section{Summary and Outlook}

We have developed a new Python-based software package - called {\sc PynPoint} - for analysing data for the direct imaging of exoplanets. This new method offers a number of improvements over what is currently available. In particular, we use a principle component analysis to model the PSF of the observations that we then use to subtract the flux coming from the central star, thus revealing the signal of a faint (planetary) companion. We have applied our technique to a set of simulations based on ADI data taken with NACO on the VLT. This dataset used an apodising phase plate, which suppresses the diffraction rings from the star on one side of the image. By inserting simulated planets into these data, we find that {\sc PynPoint} achieves a detection signal-to-noise that is 50\% larger than our {\sc LOCI} reduction at separations of $\sim$$0.59''$ and up to a factor of five better for the smaller inner working angle of $\sim$$0.29''$. We also find that when measuring the flux of the planet, {\sc PynPoint} gives more stable, and hence more accurate, results than what we find with {\sc LOCI}.

To demonstrate that {\sc PynPoint} works also on non-APP data and that it reveals real exoplanets and not only simulated ones, we have re-analysed publicly available $L'$ data from the exoplanet $\beta$ Pictoris b observed in December 2009. We clearly detect the exoplanet in our final image with a signal-to-noise close to $\sim$20. 

Our method relies on empirically constructing a basis set from the data. For this reason, it should be fairly generic and work over a range of wavelengths. The simplicity of our method also means that the number of user-specified parameters is low. Essentially, the only free parameter that needs to be chosen with care is the number of basis coefficients that are used to model the PSF. All other free parameters, such as the size of the masked region, can be automated, and their fine-tuning should not have a significant impact on the final results. 

We have written {\sc PynPoint} in an object-oriented and modular way and we plan to continuously make improvements to specific sections as we further develop and test our method. In particular, we are focused on: (i) making sure that our basis set are efficient (i.e. sparse); (ii) improving our temporal averaging over the stack; and (iii) improving our spatial algorithms for denoting. We are also in the process of testing our method on different datasets. It will be interesting to see how well {\sc PynPoint} performs at shorter NIR wavelengths, where the Strehl ratio of the observations is lower than for the data presented here. Furthermore, we hope to obtain test datasets from different instruments to investigate how {\sc PynPoint} can be optimised for individual observing programs. In so doing, we hope to soon be able to deliver a well-tested package that we can release publicly to the community, thus supporting the large, dedicated exoplanet surveys that will come online in the near future.

\section*{Acknowledgments}
AA would like to thank Roban Kramer and Eduardo Silva for useful discussions on Python and object-oriented programming.

\bibliographystyle{mn2e}
\bibliography{PynPointbib}

\label{lastpage}

\end{document}